\documentclass[onecolumn,superscriptaddress,amsmath,amssymb,floatfix]{revtex4}

\usepackage{graphicx}
\usepackage{dcolumn}
\usepackage{bm}
\usepackage{textcomp}
\usepackage{color}
\usepackage{setspace}
\usepackage{txfonts}
\begin{document}

\title{Drying paint: from micro-scale dynamics to mechanical instabilities}

\author{
Lucas Goehring$^{1,2*}$, Joaquim Li$^{1}$ and Pree-Cha Kiatkirakajorn}

\address{Max Planck Institute for Dynamics and Self-Organization (MPIDS), 37077 G\"ottingen, Germany\\ $^{2}$School of Science and Technology, Nottingham Trent University, Clifton Lane, Nottingham, NG11 8NS, UK\\ *lucas.goehring@ntu.ac.uk}

\begin{abstract}
Charged colloidal dispersions make up the basis of a broad range of industrial and commercial products, from paints to coatings and additives in cosmetics.  During drying, an initially liquid dispersion of such particles is slowly concentrated into a solid, displaying a range of mechanical instabilities in response to highly variable internal pressures.  Here we summarise the current appreciation of this process by pairing an advection-diffusion model of particle motion with a Poisson-Boltzmann cell model of inter-particle interactions, to predict the concentration gradients around a drying colloidal film.  We then test these predictions with osmotic compression experiments on colloidal silica, and small-angle x-ray scattering experiments on silica dispersions drying in Hele-Shaw cells.  Finally, we use the details of the microscopic physics at play in these dispersions to explore how two macroscopic mechanical instabilities -- shear-banding and fracture -- can be controlled.
\end{abstract}

\maketitle

\section{Introduction}
The solidification of a drying colloidal dispersion has similarities with sedimentation \cite{Buscall1987}, filtration \cite{Aimar2010}, and the freezing \cite{Peppin2006b} of multiphase fluids, as well as the solidification of polymer solutions \cite{Daubersies2012,Baldwin2011}.  A mass and momentum balance for all phases is necessary to describe the compression of the dispersed particles by a flow towards the solidification, or drying front.  The essential ideas behind models of this kind can be traced back to Kynch's theory of sedimentation \cite{Kynch1952}, or to Biot's theory of poroelasticity \cite{Biot1941}.  The former treats the evolution of a two-phase mixture with liquid-like properties, while the latter deals with flows and deformations in a mixture with solid-like properties.  In recent years, more general models have evolved that can smoothly transition between these two limiting behaviours \cite{Peppin2005,Peppin2006,Aimar2010,GoehringBook}.  

\begin{figure}[b]
\centering\includegraphics[width=135 mm]{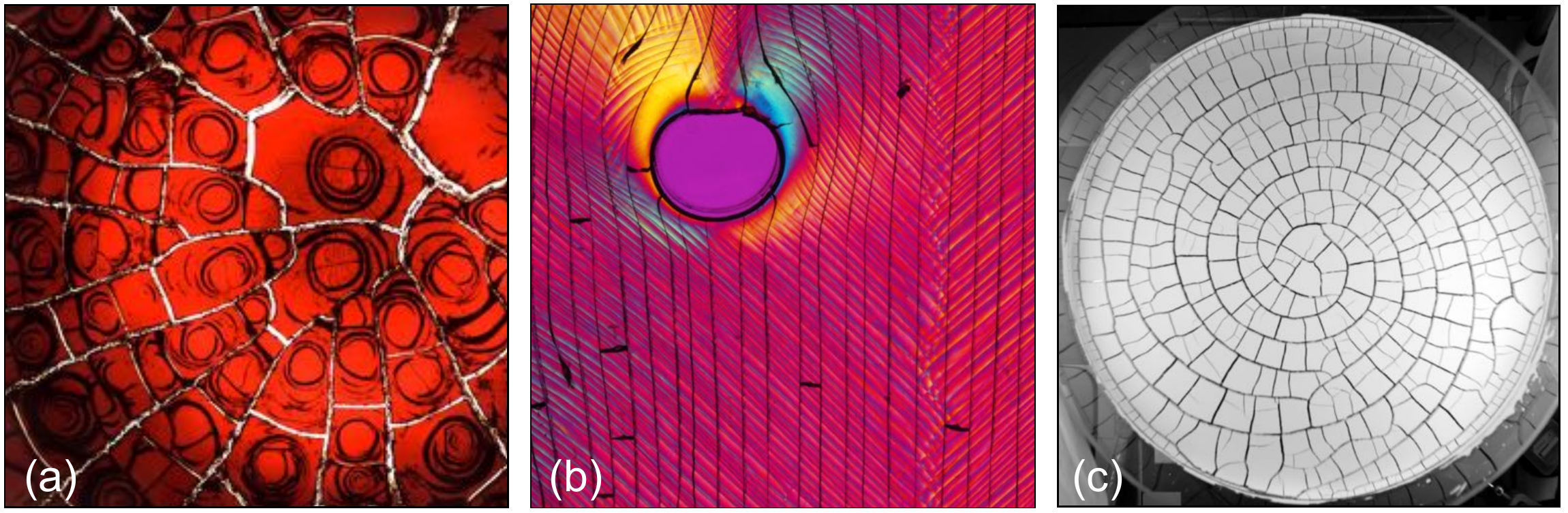}
\caption{Large-scale mechanical instabilities in dried colloidal materials can include (a) radial and spiral cracks in a blood droplet, (b) shear bands (diagonal features) and cracks (near-vertical lines) in a paint film, highlighted by viewing under crossed polarisers, and (c) the guiding of cracks in a calcium carbonate paste, that had been pre-treated by a brief rotational flow (see e.g. \cite{Nakahara2006,Nakayama2013,Kitsunezaki2016}).   Image (a) courtesy of W. Turner and D. Fairhurst, and (c) courtesy of A. Nakahara.}
\label{fig_intro}
\end{figure}

The above class of models focus on essentially mean-field, or continuum approximations of the behaviour of an enormous number of small interacting colloidal particles.  For example, in a 50 $\mu$l drop of a typical dispersion with 100 nm particles at a 10\% volume fraction, there are about one trillion particles, more than twice the number of stars in the Milky Way.  This is a comfortably large number for such approximations, yet a number of observations have shown additional effects that go well beyond the capacity of these models: the formation of crystals with rate-dependent structures \cite{Schope2006,Marin2011}, or which show fractionation and multiple-phase coexistence \cite{Cabane2016}; crystalline domains with grain boundaries that can influence flow patterns \cite{Ziane2015}; birefringence and structural anisotropy \cite{Inasawa2009,Yamaguchi2013,Boulogne2014}; plasticity during fracture \cite{Goehring2013}; and shear-banding \cite{Kiatkirakajorn2015,Yang2015}.  Some of these responses are shown in Fig. \ref{fig_intro}.  They largely involve the onset of solid-like properties, such as a yield stress or shear modulus, as a colloidal material concentrates during drying.

Here we will explore how well a continuum model of drying colloidal dispersions agrees with the behaviour of colloidal silica dispersions. These dispersions allow for a great range of colloidal interactions to be explored by varying the size of their constituent particles, and the chemistry of the dispersing fluid.  We begin by outlining a force and mass balance model, which describes the general behaviour of a drying front.  This model is completed by a Poisson-Boltzmann Cell model of inter-particle interactions, which estimates the osmotic pressure of a dispersion under different conditions.  These paired models are then tested by x-ray scattering experiments that reveal the structure of drying dispersions.  Finally, to relate this to the general theme of this special issue, \textit{patterning through instabilities in complex media}, we show how some of the more complex aspects of colloidal interactions can be used to control the shear-banding and fracture instabilities. 

\section{Theory of a 1D drying front}

We will outline here a mean-field theory of how a colloidal dispersion, such as paint, should behave when it dries under simple conditions that allow for a one-dimensional flow.  This is how a dispersion would behave in a Hele-Shaw cell (e.g. \cite{Allain1995,Dufresne2003,Dufresne2006,Daubersies2011,Giorgiutti2012,Boulogne2014}) or capillary tube \cite{Gauthier2007,Gauthier2010} that is drying from one end.  Similar models have been developed for the slightly different case of evaporation in a microfluidic microreactor, such as the designs of Salmon and collaborators \cite{Lidon2014,Ziane2015,Ziane2015b}.  The model is also compatible with the drying of polymers, such as in Refs. \cite{Daubersies2012,Baldwin2011}, if the cell model of Section \ref{S_cell} is replaced by an appropriate polymer equation of state.

\subsection{Model of a moving liquid-solid transition}

\begin{figure}
\centering\includegraphics[width=5in]{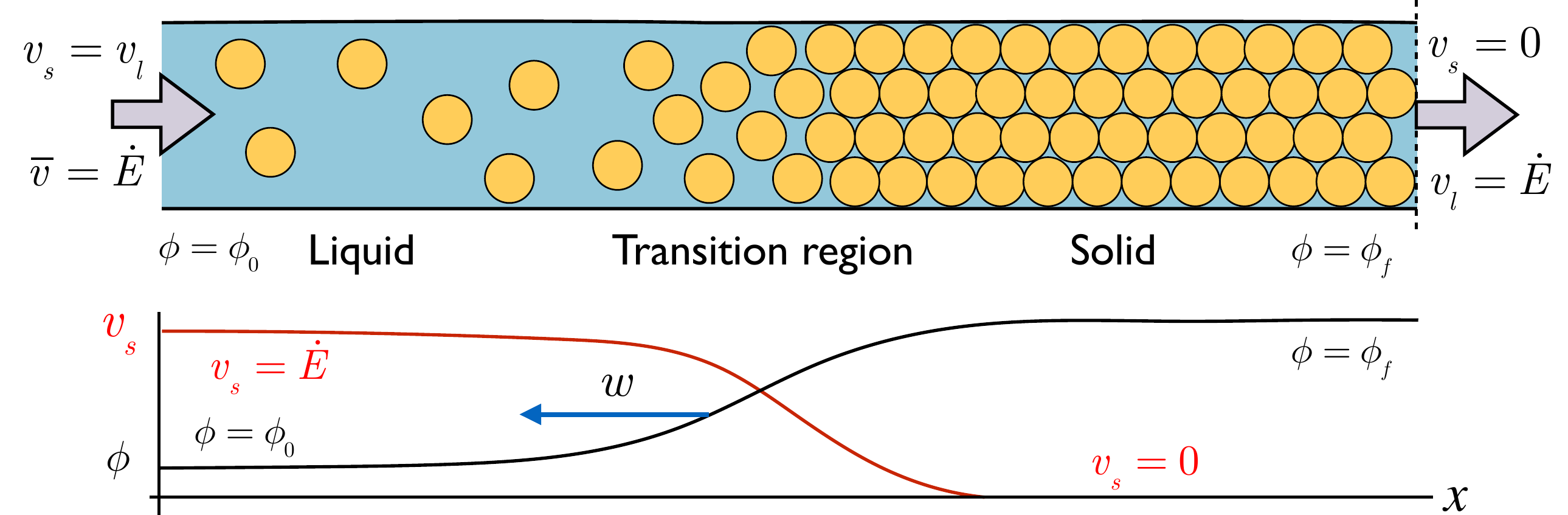}
\caption{We model the directional solidification of a drying colloidal dispersion, in one dimension.  Dispersion enters from a reservoir on the far left, at some volume fraction $\phi = \phi_0$ and evaporation occurs at the far right at a rate $\dot E$.  A solid deposit grows in from the drying edge at a velocity $w$.  Near the liquid-solid transition the particles slow down, from an initial velocity $v_s = \dot E$ to zero, and concentrate until they reach a final volume fraction $\phi_f$.   This results in a thin transition region, governed by a balance between advection and diffusion, where the properties of the dispersion change rapidly.  This drying front grows into the cell, at velocity $w$, along with the solid deposit.}
\label{model1}
\end{figure}

We consider a colloidal dispersion that is drying in a Hele-Shaw cell, as sketched in Fig. \ref{model1}.  The cell has a regular (usually rectangular) cross-section, and two ends.  Evaporation occurs at a rate (volume flux per unit area) $\dot E$ at one end, while the other end is fed by a reservoir of colloidal dispersion, with some initial volume fraction $\phi = \phi_0$.  The dispersion flows slowly through the cell, from one end to the other, along a direction $x$.  While the liquid phase can evaporate from the drying end, the solid colloidal particles must remain behind and will accumulate there.  Over time they will build up a porous solid deposit with final volume fraction $\phi = \phi_f$, that will grow back into the cell at some velocity $w$. 

Within the cell one can distinguish between the velocity of the colloidal particles, $v_s$, the velocity of the dispersant liquid, $v_l$, and a bulk velocity $\bar v = \phi v_s + (1-\phi) v_l$.   Here, all velocities are averaged over the cross-section of the cell.  This thus neglects any effects of gravity, such as sedimentation-driven instabilities near the solidification front \cite{Selva2012}.  If there are no material losses in the cell, then the total flux everywhere must balance the drying rate at the edge, $\bar{v} = \dot E$.  Far from the solidification front both the particles and the liquid will travel together at this speed.   The mean velocity, $v_s$, and volume fraction, $\phi$, of the particles can evolve, however.  In particular, they will slow and concentrate as they approach the solidification front. During this process a mass balance on the solid phase requires that
\begin{equation}
\label{eom1}
\frac{\partial \phi}{\partial t} +\frac{\partial }{\partial x} (\phi v_s)=0.
\end{equation}
We want to know how the evolution of $v_s$ and $\phi$ near the solidification front depends on the properties of the dispersion.  To do so, we now use a momentum balance to find an expression for the solid volume flux $\phi v_s$.

As the particles slow down the fluid phase must speed up, to keep the total flux across the cell constant.  The flow of water past the particles will cause drag, and transfer momentum from the fluid phase to the dispersed phase.   The pressure of a small parcel of dispersion, containing both solid and liquid, can be divided between
\begin{equation}
\label{fbal1}
P = p + \Pi
\end{equation}
where $P$ is the total, or thermodynamic, pressure; $p$ is the pressure of the fluid phase, as would be measured by a manometer through a dialysis membrane that blocks the particles \cite{Peppin2005,GoehringBook}; and $\Pi$ is the osmotic pressure of the dispersed charged particles.  Starting from the viewpoint of a compliant solid, the equivalent poroelastic balance between a total (or effective) stress $\sigma$, a stress borne by the network of particles, $\tilde\sigma$, and the fluid (or pore) pressure, $p$, can be expressed in tensor notation as
\begin{equation}
\label{fbal2}
\sigma_{ij} = -p\delta_{ij} + \tilde\sigma_{ij}
\end{equation}
where $\delta_{ij}$ is the Kroneker delta function, and the sign difference is due to the different conventions of positive stress versus pressure.  

Since there are no external forces on the dispersion, nor any body forces (we are neglecting gravity), momentum balance can be expressed as $\nabla P= 0$, or $\nabla \cdot \sigma = 0$. Considering only a one-dimensional flow along the $x$-direction, this momentum balance implies that the osmotic pressure of a fluid-like dispersion will vary according to
\begin{equation}
\label{fbal3}
\frac{\partial \Pi}{\partial x} =  - \frac{\partial p}{\partial x} =  nF_d,
\end{equation}
where $n$ is the number density of particles (i.e. for spheres of radius $a$, $n = 3\phi /4\pi a^3$), and $F_d$ is the average drag force per particle.  

For an isolated spherical particle of radius $a$ moving at a relative speed $v_s-\bar{v}$ with respect to a surrounding fluid of viscosity $\mu_0$ and average velocity $\bar{v}$, the drag force felt by the particle is the Stokes drag $-6\pi\mu_0 a(v_s-\bar{v})$.  In a dense dispersion, the hydrodynamic interactions between nearby particles will increase this drag by the factor $r(\phi)$, known variously as the hindered settling coefficient \cite{Landman1992} or the sedimentation factor \cite{Russel1989} (or a mobility, $f = 1/r$, is sometimes used \cite{Peppin2005,Peppin2006,GoehringBook}). For rigid spheres, the semi-empirical expression $r=(1-\phi)^{-6.55}$ has been suggested \cite{Russel1989,Peppin2006}, and we will adopt this here.  Colloidal interactions can be included by introducing an equation of state $\Pi = nkT Z$, where the compressibility factor $Z(\phi)$ depends on the interaction potential between particles (for example, for an ideal gas $Z=1$), and $kT$ is the Boltzmann energy (we use $T = 293$ K throughout this paper).    By combing these definitions, Eq. \ref{fbal3} becomes
\begin{equation}
kT\frac{\partial}{\partial x}(Zn) = - 6\pi \mu_0 a (v_s-\bar{v})r n.
\end{equation}
Now, by introducing the Stokes-Einstein diffusivity, $D_0 = kT/6\pi\mu_0a$, which is the diffusion constant of a single isolated spherical particle, and by using the chain rule, one can obtain the expected flux of particles past any point as
\begin{equation}
\phi v_s =  \phi \bar{v} - \bigg(\frac{D_0}{r(\phi)}\frac{\partial \phi Z}{\partial \phi}\bigg) \frac{\partial \phi}{\partial x}.
\end{equation}
Here, the two terms on the right hand side of the equation correspond to the advective flux along the cell, and the diffusive flux down any concentration gradients, respectively.   The latter can be simplified by collecting all the inter-particle interactions into a dimensionless diffusivity 
\begin{equation}
\label{eqD}
\tilde{D}(\phi) = \frac{1}{r(\phi)}\frac{\partial \phi Z}{\partial \phi},
\end{equation}
such that
\begin{equation}
\phi v_s = \phi \bar{v} - D_0 \tilde{D} \frac{\partial \phi}{\partial x}.
\end{equation}
In other words, for any concentration (or collective) diffusivity $D$, the dimensionless diffusivity $\tilde{D} = D/D_0$.  

Introducing the above particle flux into the mass balance of Eq. \ref{eom1}, one obtains the usual (e.g. \cite{Peppin2005,Peppin2006,Daubersies2011,Boulogne2014}) advection-diffusion model of colloidal transport,
\begin{equation}
\frac{\partial \phi}{\partial t} +\frac{\partial}{\partial x} \bigg(\phi \bar{v} - D_0 \tilde{D} \frac{\partial \phi}{\partial x}\bigg) = 0.
\end{equation}
Here, given a model for $\tilde{D}$, developed in the next section, we look for a steady-state solution that describes the jump in concentration associated with the liquid-solid transition, in a reference frame that is co-moving with the drying front.  If the front is growing into the cell at a fixed velocity $w$, then the transformation $x^\prime = x - wt$ introduces an additional advection term, giving
\begin{equation}
\label{comoving}
\frac{\partial \phi}{\partial t} +\frac{\partial}{\partial x^\prime} \bigg(\phi (\bar{v}-w) - D_0 \tilde{D} \frac{\partial \phi}{\partial x^\prime}\bigg) = 0.
\end{equation}
If we seek a steady-state solution, then the term inside the spatial derivative of Eq. \ref{comoving} must be a constant.  In the reservoir (i.e. in the limit of $x^\prime \rightarrow -\infty$) we have the boundary condition $\phi = \phi_0$, allowing one to write down a simple first-order equation describing the evolution of the volume fraction across the liquid-solid transition,
\begin{equation}
\label{phiODE}
\frac{\partial \phi}{\partial x}  = \frac{(\bar{v}-w)(\phi - \phi_0)}{D_0 \tilde D}  = \frac{ (\phi - \phi_0) }{L \tilde{D} },
\end{equation}
where $L = D_0 /(\bar{v}-w)$ sets the natural length-scale of the front, and the effects of all particle interactions are contained in $\tilde{D}$.  To solve this we choose some arbitrary value for $\phi$ at the origin, typically 0.3, and use Matlab's non-linear ODE solver to numerically integrate Eq. \ref{phiODE} in both directions. Such solutions will form the basis of comparison to experiments in Section \ref{SAXS}.  Performing the same transformation directly on the mass balance of Eq. \ref{eom1}, and looking for the co-moving steady-state solution, then allows us to simultaneously solve for the particle velocity via
\begin{equation}
\frac{v_s}{\bar{v}-w} = \frac{\phi_0}{\phi}-\frac{\phi_0}{\phi_f}.
\end{equation}

Finally, it is interesting to note that much of the above discussion can also be applied when the dispersion is concentrated to the point of behaving as a porous solid \cite{Peppin2005,GoehringBook}.  In this context the compressibility factor $Z$ can be related to the relevant bulk modulus of the collective assembly, or network, of particles by
\begin{equation}
K = \phi \bigg(\frac{\partial \Pi}{\partial \phi}\bigg)_{T,N,p} = nkT\frac{\partial \phi Z}{\partial \phi},
\end{equation}
where the derivative is taken at a constant temperature $T$ and number $N$ of particles, and fluid pressure $p$.  In poroelasticity this is often referred to as the drained bulk modulus \cite{GoehringBook}, as its definition allows for exchange of water molecules with some reservoir (as through a dialysis sack).  A second elastic modulus, such as a shear modulus or Poisson's ratio is, however, required to complete any mechanical description of a solid.  We are not aware of any model that attempts to predict such additional moduli in the context of colloidal materials.

\subsection{Osmotic pressure and the Poisson-Boltzman cell model}
\label{S_cell}

To evaluate the advection-diffusion model described above we need its equation of state, which takes the form of an expression for the osmotic pressure, $\Pi(\phi)$, or compressibility factor $Z(\phi)$.  When charged particles are dispersed in an electrolyte solution they affect the distribution of ions in the solution. The osmotic pressure of the dispersion can include contributions from the hard-sphere interactions of the particles ($\Pi_{\textrm{s}}$), and from the deformation of the clouds of ions around each particle, as the particles are concentrated ($\Pi_{\textrm{q}}$). We thus assume a function of the form
\begin{equation}
\label{p_osm}
\Pi = \Pi_{\textrm{s}} + \Pi_{\textrm{q}} = nkT(Z_s+Z_q)
\end{equation}
Of course, additional contributions to the osmotic pressure could also be considered.  The models presented in Ref. \cite{Bonnet-Gonnet1994,Gromer2015} discuss how to add an attractive van der Waals term to $\Pi$, for example, before neglecting it as only relevant for very small particle separations.  

For the entropic term we use the Carnahan and Starling \cite{Carnahan1969} equation of state for hard spheres,
\begin{equation}
\label{CS}
Z_{s} = \frac{1+\phi + \phi^2-\phi^3}{(1-\phi)^3}.
\end{equation}
Although there are known to be observable deviations from this at high volume fractions ($\phi \gtrsim 0.55$ \cite{Peppin2006}), we do not wish to introduce a divergence in $Z$ at close packing \cite{Hall1972,Gromer2015} or random close-packing \cite{Peppin2006,Daubersies2011} that are included in other approximations of the hard-sphere equation of state.  Rather, as in Ref. \cite{Goehring2010}, for a slowly increasing $\phi$ we would expect van der Waals attraction to cause irreversible aggregation, before any such divergence would be physically relevant.

\begin{figure}[]
\centering\includegraphics[width=135 mm]{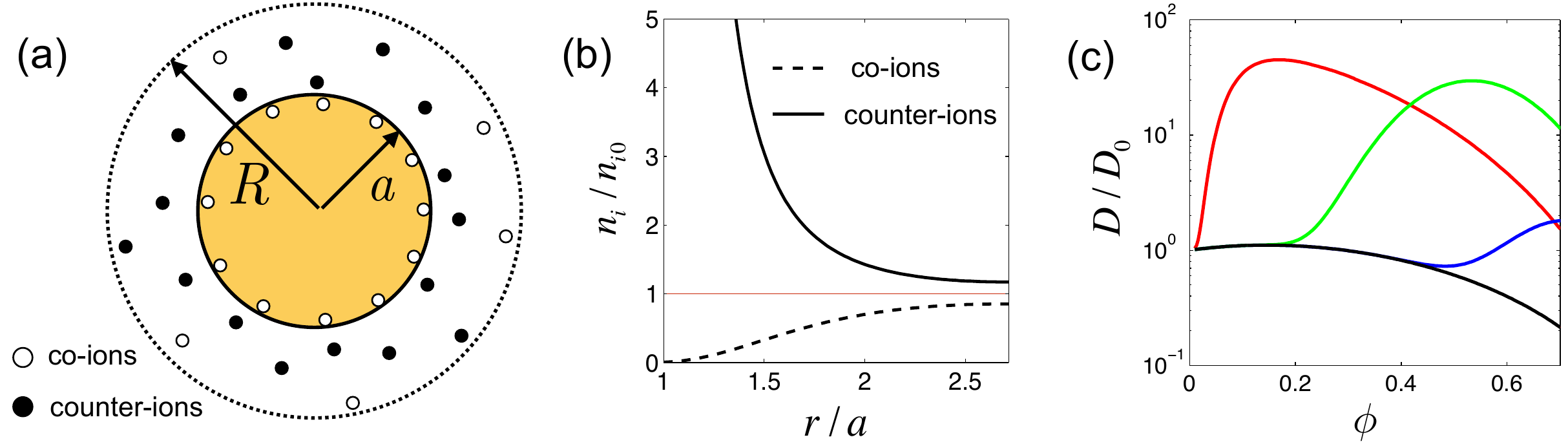}
\caption{Poisson-Boltzmann Cell (PBC) model.  (a) An electrically-neutral spherical cell is constructed around each charged particle.  (b) Within this cell the equilibrium distribution of ions is shown for the example case of colloidal silica with radius $a = 8$ nm and surface charge density $\sigma$ = 0.5 $e$/nm$^2$ at volume fraction $\phi = 0.05$ in equilibrium with a monovalent salt of concentration $n_0 = 5$ m$M$. The ionic concentrations at the outer surface of the cell are related to the osmotic pressure of the dispersion.  (c) These colloidal interactions will typically increase the effective diffusivity of the particles, $D/D_0$, above that of hard spheres (black curve, $Z_q =0$).  Peaking from left to right are shown are the cases for (red) $a = 8$ nm,  $\sigma$ = 0.5 $e$/nm$^2$ and $n_0 = 5$ m$M$, (green) $a = 99$ nm,  $\sigma$ = 0.033 $e$/nm$^2$ and $n_0 = 1$ m$M$ and (blue) $a = 99$ nm,  $\sigma$ = 0.033 $e$/nm$^2$ and $n_0 = 5$ m$M$.}
\label{fig_PB}
\end{figure}

To describe the electrostatic contribution to the equation of state we evaluate the effective pair-potential between nearby particles through the Poisson-Boltzmann Cell (PBC) model  \cite{Alexander1984,Belloni1998,Trizac2003,Jonsson2011}, which solves the fully non-linear Poisson-Boltzmann equation on electrically neutral domains around each particle.   This model divides the total volume of a small parcel of dispersion up equally between all the particles within it, assigning a spherical `cell' of radius $R = a/\phi^{1/3}$ to each particle, as sketched in Fig. \ref{fig_PB}(a).  In other words, each particle lives in its own neutrally-charged sphere, which shrinks as the volume fraction increases.  Inside any cell one solves the Poisson-Boltzmann equation that describes the interactions of an equilibrium distribution of ions and the electric field that they generate.  Generally, this takes the form \cite{Russel1989}
\begin{equation}
\epsilon\epsilon_0\nabla^2\psi = -e\sum_{i=1}^Nz_{i}n_{i0}e^{-ez_i\psi/kT}
\label{PB1}
\end{equation}
where $\epsilon_0$ is the permittivity of free space, $\epsilon$ is the dielectric constant of the fluid, $\psi$ is the electrostatic potential field, $e$ is the fundamental charge, $kT$ is the thermal energy and $z_i$ is the relative charge of chemical species $i$ with some background number density $n_{i0}$ (defined by the electrolyte concentration when $\psi=0$).  

We consider a monovalent electrolyte of equilibrium concentration $n_0$.  For a colloidal dispersion that has been dialysed, this will be the concentration of the electrolyte in a solution that is in Donnan equilibrium with the dispersion across the dialysis membrane \cite{Belloni1998,Trizac2003}.  For a monovalent electrolyte Eq. \ref{PB1} simplifies to
\begin{equation}
\label{PB2}
\nabla^2\varphi = \kappa^2\sinh{\varphi}
\end{equation}
where 
\begin{equation}
\label{redpot}
\varphi = \frac{e\psi}{kT}
\end{equation}
is the reduced electrostatic potential, and $\kappa^{-1}$ is the Debye length, defined through
\begin{equation}
\label{debye}
\kappa^2 = \frac{2e^2n_0}{\epsilon\epsilon_0kT}.
\end{equation}
The distributions of positive (+) and negative (-) ions in the cell can then be given by $n_\pm = n_0e^{\mp\varphi}$, and a typical ion distribution is shown in Fig. \ref{fig_PB}(b).

As boundary conditions for Eq. \ref{PB2}, Gauss' law is used to equate the electric flux out of a sphere to the charge contained within it.  Since the total cell is charge-neutral, this means that $\partial\varphi/\partial r = 0$ at $r = R$.  The surface of the particle is charged, with a surface charge density $\sigma$.  This gives $\partial \varphi / \partial r = -4\pi L_B \sigma$ at $r = a$, where $L_B = e^2/4\pi\epsilon\epsilon_0kT$ is the Bjerrum length (0.7 nm in water at room temperature).

We are interested in the osmotic pressure of the dispersion.  In equilibrium, this pressure must be constant throughout the cell, and is easily calculated at the outer surface of the cell, where the electric field vanishes.    The osmotic pressure is then simply the difference between the chemical potential of the ions there and in an electrically neutral solution of ions at concentration $n_0$ \cite{Alexander1984,Bonnet-Gonnet1994,Jonsson2011}.  In other words,
\begin{equation}
\label{p_PB}
\frac{\Pi_{q}}{kT} = (n_+(R)-n_0) + (n_-(R)-n_0) = n_0(e^{-\varphi(R)}+e^{\varphi(R)}-2).
\end{equation}

The PBC model, as described above, has been used to model the osmotic pressure of a range of colloidal materials, including colloidal polystyrene \cite{Bonnet-Gonnet1994,Reus1997} and silica \cite{Jonsson2011,Li2015,Cabane2016} under a variety of conditions.  It can also be used to estimate the effective pair-potential of the particles, predicting an effective (or renormalized) surface charge, surface potential, and Debye length, for example \cite{Alexander1984,Trizac2003}.  We implemented the PBC model in Matlab, and checked the code against results in Refs. \cite{Belloni1998,Jonsson2011}, which it was able to reproduce.  The osmotic pressures from it were then used to calculate the concentration diffusivity of a dispersion via Eq. \ref{eqD}, given its bare surface charge density $\sigma$, radius $a$, and the equilibrium salt concentration $n_0$.   A typical prediction of $D(\phi)$ is shown in Fig. \ref{fig_PB}(c), and the code underlying the cell model is provided as online supplemental material.  

\section{Drying fronts observed by x-ray scattering}
\label{SAXS}

Small angle x-ray scattering (SAXS) was used to observe the drying fronts in colloidal silica of different particle sizes and salt concentrations.  Our primary aim was to evaluate the accuracy of the combined advection-diffusion and Poisson-Boltzmann cell models in describing the concentration gradients across the drying front, and to identify when additional effects were important.  Additionally, the sample preparation for these experiments allowed us to make precise tests of the Poisson-Boltzman cell method of predicting the osmotic pressure of strongly charged colloidal particles.

\subsection{Materials and methods}
\label{SAXSmethods}

\begin{figure}[]
\centering\includegraphics[width=135 mm]{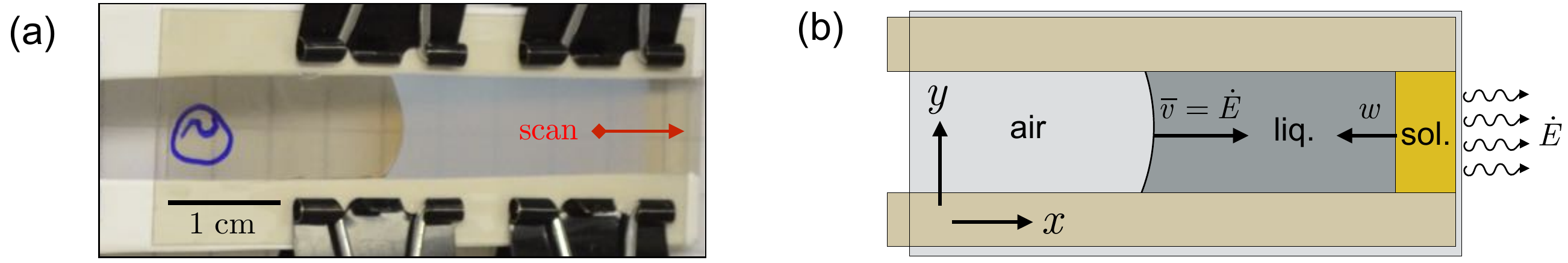}
\caption{Small-angle x-ray scattering (SAXS) sample geometry.  (a) Dispersions were prepared by drying in narrow Hele-Shaw cells. One end of the cell was raised by a few mm, to allow the dispersion (cloudy intermediate region) to drain to the other edge, where evaporation occurred. A solid deposit (clear, right-hand side) then slowly grew in from this edge.  After several hours cells were raised vertically, and SAXS spectra were collected at a series of scan points along the mid-line of the cell. (b) The sample geometry is idealised as a receding meniscus moving at a velocity $\bar{v} = \dot{E}$, to balance evaporation at rate $\dot{E}$ from one end, and a solid deposit growing in the opposite direction at velocity $w$.}
\label{saxs_methods}
\end{figure}

Aqueous dispersions of colloidal silica (Ludox SM 30, HS 40 and TM 50, Sigma-Aldrich) were passed through a 5 $\mu$m teflon filter, and then cleaned by dialysis for two days against an aqueous solution of NaCl (concentrations between 0.5 - 50 m$M$) and NaOH (0.1 m$M$, to bring the measured pH to 10).  The washed dispersions were then compressed by the osmotic stress method, as detailed in \cite{Jonsson2011,Li2015}.  Briefly, the concentration was slowly raised by the addition of polyethylene glycol (PEG 35000, Sigma-Aldrich) to the bath on the outside of the dialysis sack, while keeping the NaCl and NaOH concentrations there fixed.  This solution was changed every two days, for a period of at least six days.  The polymer could not pass through the dialysis sack (molecular cutoff 14 kD), and so created a pressure difference that was balanced when the osmotic pressure of the colloidal dispersion in the sack reached an equilibrium concentration.  After dialysis the volume fraction of each dispersion was determined through weighing a small sample before, and after, drying at 120$^\circ$C in an oven overnight.  For this calculation we assume a silica density of 2200$\pm$50 kg/m$^3$ \cite{Bergna2005,Iler1979,Cabane2016}.

Hele-Shaw cells were constructed out of two 26$\times$52 mm$^2$ mica sheets, 35-50 $\mu$m thick, as sketched in Fig. \ref{saxs_methods}.  A pair of 0.3 mm thick plastic spacers, wrapped in teflon tape, were placed between the mica sheets, along the long edges of each cell. This created a space about 1 cm wide, and 5 cm long, which was open on both short ends.  These cells were clipped together, filled approximately half-full of dispersion, and then left to dry for $\sim$ 8 hours.  During this time one open end was raised slightly to allow the dispersion to settle to the other side, from which evaporation proceeded at a rate $\dot{E}$.  The large air-gap to the other edge rendered negligible any evaporation from the other open side of the cell. Time-lapse images were then taken of the cells as they dried, at intervals of ten minutes, and the evaporation rate was measured by tracking the velocity $\bar{v}$ of the retreating meniscus in the cell, on the assumption that $\bar{v} = \dot{E}$.  The velocity at which the solid phase grew back into the cell, $w$, was also directly measured through the image sequence.  In all cases the relative speed of the particles with respect to the drying front, $\bar{v} - w$, was between 0.36 and 0.58 $\mu$m/s, with an estimated error on each measurement of about 10\%.   

After about 8 hours of drying the cells were raised vertically and placed in the path of an x-ray beam.  The small-angle x-ray scattering experiments were performed with beamline ID02 at the ESRF at an energy of 12.4 keV, using detector distances of 2.5 and 10 m.  An elliptical beam was used, characterised by a full-width half-maximum of intensity of 50-70 $\mu$m in the vertical direction, and 250-400 $\mu$m in the horizontal.  As in \cite{Boulogne2014} the structure of the colloidal dispersions near the liquid-solid transition was characterised by moving the sample vertically across the path of the beam, in periodic steps of between 60-200 $\mu$m.  Typically 50 to 100 spectra were collected along each scan line, over a period of about 5 minutes.  

\subsection{Results}

\begin{figure}[]
\centering\includegraphics[width=135 mm]{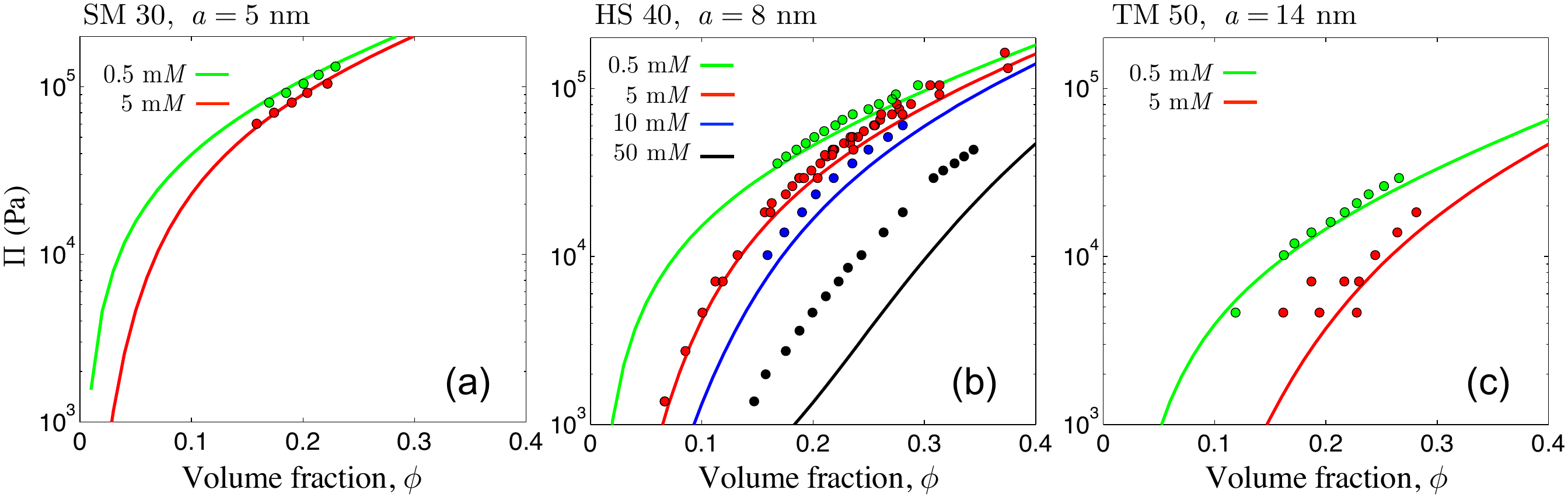}
\caption{Osmotic compression experiments on colloidal (a) SM 30 (particle size $a$ = 5 nm) (b) HS 40 ($a = 8$ nm) and (c) TM 50 ($a = 14$ nm) show how the osmotic pressure of the silica dispersions depends on their volume fraction, particle size, and the salt concentration outside the dialysis sack.  The solid lines give predictions from the Poisson-Boltzmann cell model (Eq. \ref{p_osm}), under corresponding conditions, assuming a bare surface charge density of $\sigma = 0.5$ $e$/nm$^2$.  Some data for HS 40 at 5 m$M$ are incorporated from Ref. \cite{Li2015}.  All figure data are provided as online supplementary information.}
\label{fig_pressure}
\end{figure}

In preparation for our scattering experiments we dialysed about a hundred samples of colloidal silica against standard solutions of PEG.  This provided dispersions with a range of particle sizes, salt concentrations and solid volume fractions.  For each sample the osmotic pressure was determined from the equation of state for PEG given in \cite{Li2015}, as in \cite{Jonsson2011,Li2015,Cabane2016}.  Then, the Poisson-Boltzman Cell (PBC) model was used to predict the corresponding osmotic pressures, by Eqs. \ref{p_osm}, \ref{CS} and \ref{p_PB}.  For this calculation a bare surface charge density of 0.5 $e$/nm$^2$ was assumed, based on titration \cite{Bolt1957,Persello2000,Jonsson2011}.  

The results of these osmotic compression tests, shown in Fig. \ref{fig_pressure} and tabulated in the online supplemental information, demonstrate the excellent agreement of the PBC model with the experimental equation of state of the colloids at low-to-intermediate salt concentrations, and intermediate colloid concentrations (when dispersions still behave rheologically as a liquid).  For example, for 0.5 m$M$ NaCl the PBC model accurately predicts the osmotic pressure of all three types of dispersions to within 15\%, with no free parameters.  Agreement becomes less good for higher salt concentrations, however, until the PBC model systematically under-predicts the osmotic pressure of HS 40 at 50 m$M$ NaCl by about a factor of two, in the range of concentrations studied.  Nevertheless, given the complexity of the interactions between the densely packed, highly charged colloidal particles, this level of agreement is still satisfying.

\begin{figure}[]
\centering\includegraphics[width=135 mm]{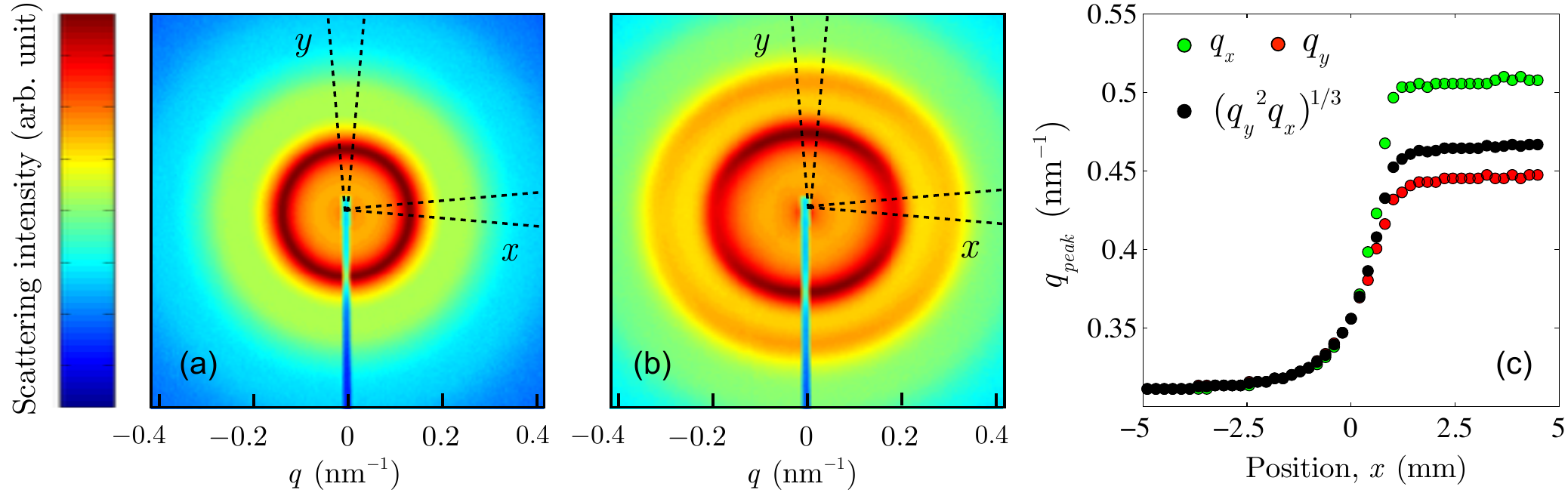}
\caption{Scattering spectra (a,b) were collected from locations across the liquid-solid transition of dispersions drying in Hele-Shaw cells.  From each two-dimensional spectrum we extracted the scattering intensity along the $x$ and $y$ directions by azimuthally averaging over arcs $\pm 5^\circ$ to the respective axis.  For low volume fractions (a) the dispersions all behave as liquids, with no anisotropy between the $x$ and $y$ directions.  After some critical volume fraction (here, about $\phi = 0.38$), the structure factors along the two directions begin to differ: the particles start to pack closer together along the $x$-axis than in the other directions.  (c) From the structure factors we find the $q$-values corresponding to the maximum of the primary scattering peak along the $x$ ($q_x$) and $y$ ($q_y$) directions, from which we calculate the volume fraction $\phi$ and strain $\gamma$.}
\label{saxs_spectra}
\end{figure}

Five samples were used for drying experiments in Hele-Shaw cells, where we extracted volume fraction profiles from series of x-ray spectra collected across the drying fronts.  These cells contained dispersions of the three types of colloidal silica with either 0.5 or 5 m$M$ NaCl, and initial volume fractions of about 0.2.  The resulting spectra were analysed as in Ref. \cite{Boulogne2014}, which focussed on the onset of anisotropy and birefringence in a similar experiment.  Briefly, as shown in Fig. \ref{saxs_spectra}, from each spectrum we measured the position of the main scattering peak of the structure factor (obtained by dividing the scattering intensity by a form factor of dilute particles) in two orientations: $q_x$ parallel to the flow through the channel, and $q_y$ perpendicular to it. For the third direction we assume that $q_z = q_y$, as the dispersion is being compressed only along the $x$-axis (n.b. this assumption was tested in Ref \cite{Boulogne2014}).  From these results we then calculated the volume fraction 
\begin{equation}
\label{phi_calc}
\phi = c(q_xq_y^2)
\end{equation}
and a deviatoric strain 
\begin{equation}
\label{gamma_calc}
\gamma = \frac{2}{3}\bigg(\frac{q_x}{q_y} -1\bigg),
\end{equation}
as derived in \cite{Boulogne2014}. In Eq. \ref{phi_calc} the constant of proportionality, $c$, was found for each type of dispersion by measuring the position, $q_p$, of the main scattering peak in each of the calibration samples that were used in the osmotic stress test, and fitting them to $\phi = c q_p^3$, as in \cite{Li2012}.  The volume fraction tracks the volumetric strain (or compression) of the system as it dries, while $\gamma$ characterises any volume-preserving, but shape-changing strains, such as can result from shears.  For a liquid-like response one expects that $\gamma = 0$, as $q_x = q_y$.  

\begin{figure}[]
\centering\includegraphics[width=135 mm]{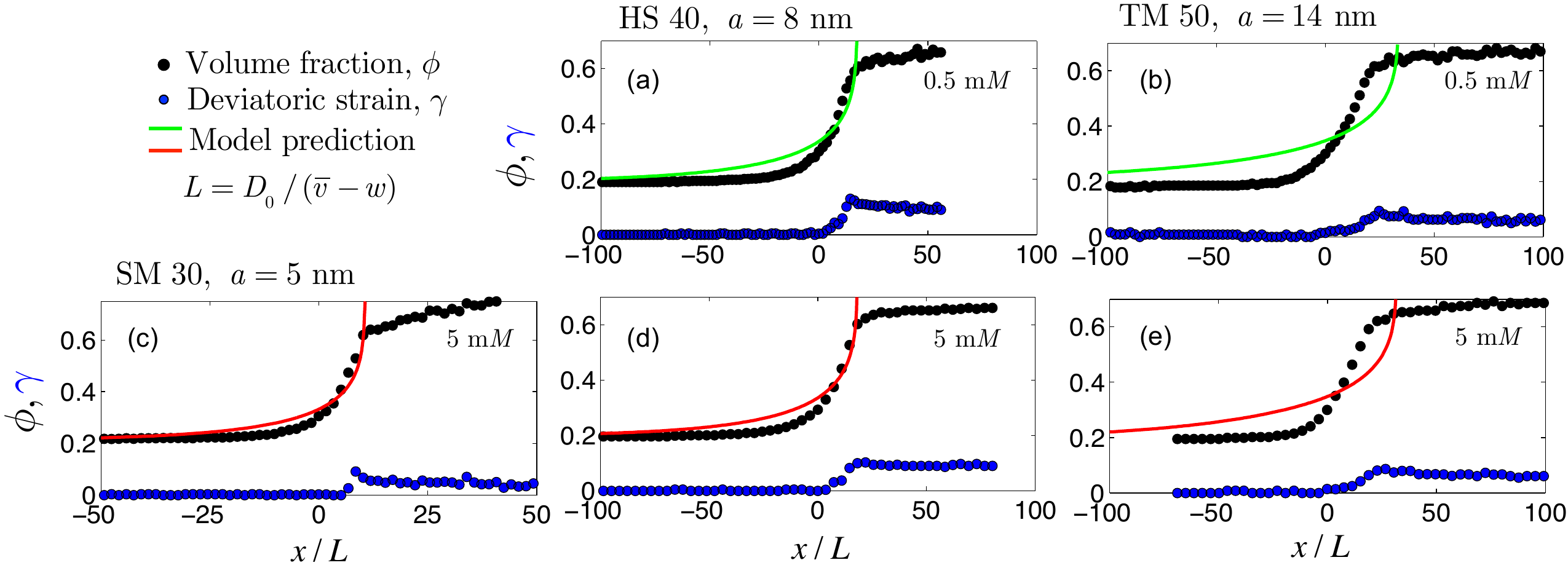}
\caption{The liquid-solid transition was observed in the drying of five different dispersions, with various particle sizes and salt concentrations.  The solid volume fraction slowly increases from the left to the right of each graph, as the particles are packed closer and closer to each other.  Distances are rescaled by the advection-diffusion length $L = D_0/(\bar{v} - w)$, to highlight the effects of charge on stretching out the transition region (for non-interacting particles, the transition should occur over a length $L$).  Shown are the volume fraction ($\phi$, black points, Eq. \ref{phi_calc}) and deviatoric strain ($\gamma$, blue points, Eq. \ref{gamma_calc}).  The results of the model calculations are shown as solid curves, assuming a surface charge density $\sigma = 0.5$ $e$/nm$^2$. All figure data are provided as supplemental online information.}
\label{saxs_scaled}
\end{figure}

Figure \ref{saxs_scaled} shows the results of these experiments, matched with the corresponding model predictions.  All data, including the unscaled observations, are provided as online supplemental information. For the experimental data the origin of the $x$-axis was arbitrarily centred where $\phi=0.3$.  In the model the initial volume fraction $\phi_0$ was taken to be the smallest recorded experimental $\phi$, and the origin of the $x$-axes was positioned by hand such that the model drying curves coincided with the data as well as possible.  Otherwise, there were no free parameters in the model, which uses the same particle properties as in Fig. \ref{fig_pressure}; in other words, the salt concentration and particle size are set by the corresponding experimental dispersion, and we assume a particle surface charge of $\sigma = 0.5$ $e$/nm$^2$. 

In both experiment and model the particle volume fraction rises characteristically as one crosses the liquid-solid transition.  There is then a kink in the experimental compression curves after the particles aggregate \cite{Li2012,Boulogne2014}, followed by a much more gradual compression of the solid phase in response to the large capillary pressures that occur there.  Qualitatively, these trends match the type of compression curves that have been seen in other drying droplets of complex fluids \cite{Daubersies2011,Li2012,Daubersies2012,Giorgiutti2012,Boulogne2014,Ziane2015}.  Additionally, the drying dispersions all become anisotropic (i.e. $q_x \neq q_y$) after some critical volume fraction between $\phi = 0.33$ (for the TM 50 at 0.5 mM) and $0.47$ (for the SM 30).  As in Ref. \cite{Boulogne2014} the deviatoric strain then rapidly accumulates in the dispersion, reaching a maximum of about 0.1 by the end of the liquid-solid transition.  This strain then decreases slightly in the solid region, as cracks form to release the total stress in the film. 

In all our experiments the transition from a liquid-like dispersion to an aggregated solid film extends over about 1-2 mm in real space.  Rescaled by the advection-diffusion length, $L = D_0/(\bar{v} - w)$, we can observe exactly how inter-particle interactions affect this compression of the dispersion, during drying.  Point-like particles, behaving like an ideal gas, would lead to a relatively sharp drying front where $\phi - \phi_0 \sim e^{x/L}$.   The high charge of the silica particles causes strong electrostatic interactions, which increases the width of the solid-liquid transition by a factor of about ten above the non-interacting case.  In particular, the fronts remain surprisingly well fit by a simple exponential increase in concentration, but where the exponential behaviour ranges from a characteristic length of $6.6L$ for the smaller SM 30, to $15.6 L$ for the larger TM 50.  This effect is captured by the advection-diffusion model, but the model somewhat over-estimates the width of the front, in all cases; this is particularly apparent for the TM 50 dispersions.  

\begin{figure}[]
\centering\includegraphics[width=135 mm]{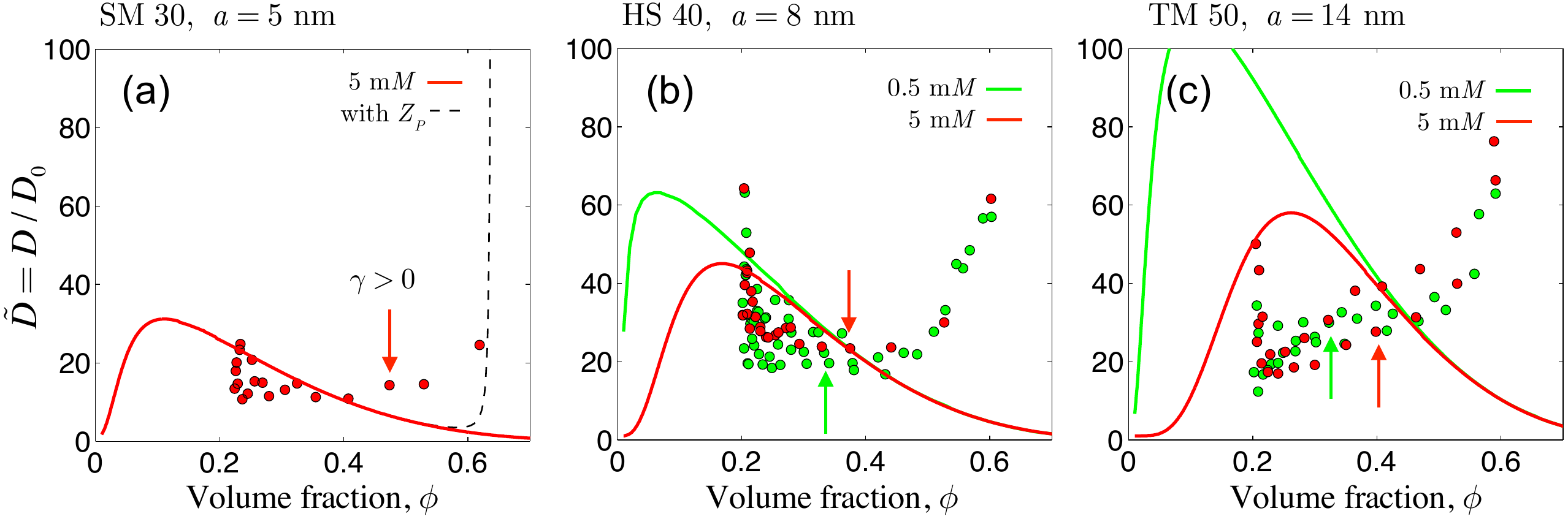}
\caption{Measured and predicted dimensionless diffusivities for colloidal silica in Ludox (a) SM 30, (b) HS 40, and (c) TM 50, at equilibrium salt concentrations of 5 m$M$ (red) and 0.5 m$M$ (green).  The large experimental scatter near $\phi = 0.2$ is an artefact of numerical differentiation of slowly varying data.  Arrows show the first volume fraction where the deviatoric strain, or anisotropy, is measurably non-zero.  In other words, it marks the point where the dispersion begins to act as a solid, with a yield stress.  In (a) we also show the corresponding prediction (as a dashed line) where Eq. \ref{CS} is replaced by the modified hard-sphere compressibility factor $Z_P$ from \cite{Peppin2006}, which diverges at random close packing. Figure data are provided as supplemental online information.}
\label{saxs_diff}
\end{figure}

To look more carefully at the physics of the liquid-solid transition we took a numerical derivative of the data in Fig. \ref{saxs_scaled}, and used Eq. \ref{phiODE} to estimate the dimensionless diffusivity $\tilde{D}$ from each experiment.  The results of this process are shown in Fig. \ref{saxs_diff}.  Generally, for both experiment and model, the larger the charged particle, the more the effective diffusivity is enhanced by the inter-particle interactions.  For the SM 30 and HS 40 dispersions, the model diffusivity agrees reasonably well with the experimental data at intermediate volume fractions, namely in the range from 0.2--0.4.   Above this they differ noticeably: the model predicts a decreasing diffusivity, approaching that of hard spheres (see also e.g. Fig. \ref{fig_PB}), whereas the data turn distinctly upwards.  For TM 50 the model and experiment substantially disagree, although the observed diffusivity shows the same trends as the other experiments, increasing quickly at high $\phi$.  These differences could suggest additional, or non-DLVO, interactions. 

The modified Carnahan-Starling equation proposed by Peppin, Elliot, and Worster (Ref. \cite{Peppin2006} Eq. 17; matched asymptotic solution between Carnahan-Starling at low $\phi$, and molecular dynamics simulations at high $\phi$) does also turn back upwards at large volume fractions, and in fact diverges near random close packing ($\phi = 0.64$).  However, if this compressibility factor, $Z_P$, is used in place of Eq. \ref{CS}, there is no noticeable difference in the response below about $\phi = 0.60$, as demonstrated by the dashed line in Fig. \ref{saxs_diff}(a).  It cannot account for the observed increase in the effective diffusivity of the colloidal particles in the range of $\phi = 0.4$ to 0.6.
 
Instead, the increase in the collective diffusivity of the particles appears to be associated with the onset of a macroscopic yield stress of the dispersions.  On Fig. \ref{saxs_diff} we also indicate the volume fractions corresponding to the first detection of structural anisotropy in our dispersions (i.e. the first values of $\phi$ where $\gamma$ is noticeably non-zero, in Fig. \ref{saxs_scaled}).   These concentrations mark the point where the dispersions acquire both a yield stress, and a finite shear modulus, and where the individual particles will start being caged by strong interactions with their neighbours.  The unexpected increase in $\tilde{D}$ occurs at, or shortly after, the particles start to behave together as a weak, soft solid. 

\section{Patterns and instabilities driven by drying fronts}

The previous sections have explored first a simple theoretical description of a drying front in a colloidal dispersion, then an experimental investigation of such fronts via x-ray scattering techniques.  We showed that a force and mass balance allowed us to predict how a colloidal dispersion is compressed along one axis as it dries, and how this compression affects the state of the dispersion.  Here we will attempt make connections between these results and the macroscopic mechanical instabilities that accompany drying, namely the appearance of shear bands and cracks in a drying colloidal film.  In particular, we will show that the magnitude of shear relieved by the shear bands is controlled by the total amount of strain accumulated across the liquid-solid transition, and that the anisotropy caused by the transition can also control the paths of any subsequent cracks that form.

\subsection{Control of shear bands}

\begin{figure}[]
\centering\includegraphics[width=135 mm]{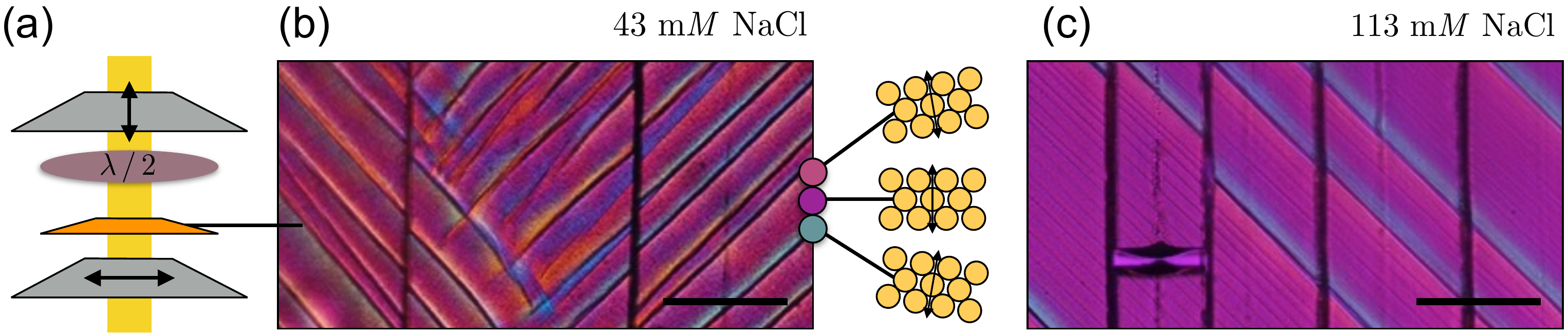}
\caption{Shear bands in a colloidal film can be visualised by (a) polarisation microscopy of the film between crossed polarisers, and a half-wavelength filter.  (b) Under these conditions, and white light illumination, the film will appear purple when the optic axis is aligned with the initial polariser.  Birefringence is visible by colour changes.  As shown here, clockwise rotation of the optic axis shifts the colour to blue, while counterclockwise to red.  (c) The addition of salt to the dispersion reduces the intensity of these colour variations, i.e. the total shear felt across the shear bands.  In some cases, for very high salt concentrations, additional fainter shear bands can be seen between the main lines.  Scale bars are 200 $\mu$m.}
\label{shearbands}
\end{figure}

Dried colloidal films frequently show regular bands or strips, arranged in a chevron pattern, as in Fig. \ref{shearbands}.  Although such features have often been noticed (e.g. \cite{Hull1999,Berteloot2012,Boulogne2014}), they have only recently been shown to be shear bands \cite{Kiatkirakajorn2015,Yang2015}, and they form at $\pm45^\circ$ to the direction of drying.  Here we will show how the amount of slip (or the magnitude of the shear) accommodated by these bands is controlled by the liquid-solid transition. In other words, we will demonstrate that the shear-band instability can be manipulated through changes to the chemistry of a drying dispersion.  
 
We explored shear bands in a Hele-Shaw geometry, using silica dispersions; the experiments are similar to those described in Section \ref{SAXS}.   The distortion around the bands can be visualised, and quantified, by polarisation microscopy, as in Ref. \cite{Kiatkirakajorn2015}.  Briefly, dried drops or films of dispersions are generally birefringent \cite{Inasawa2009,Yamaguchi2013,Boulogne2014,Kiatkirakajorn2015}, as their material has been compressed along the direction of flow, during drying \cite{Boulogne2014}.  The direction of compression defines, on average, the optic axis of the film.   Light with a polarisation that is either parallel or perpendicular to this axis will pass through the film unchanged -- all other polarisations will be modified by the film.   The shear bands focus distortion into their immediate neighbourhood, and thus can locally reorient the optic axis, rotating it one way or the other.   If the sample is between crossed polarisers, this rotation is visible as a change in the colour and/or intensity of transmitted light.

Of the types of colloidal silica used in this study, Levasil 30 (AkzoNobel; particle radius $a=46$ nm) has a particularly strong birefringence.  As received, it is dispersed in a solution of $\sim$ 27 m$M$ NaCl (measured by conductivity measurements of the supernatant liquid after centrifugation; ions and approximate value confirmed by manufacturer).  To test how salt changes the shear bands, samples of this dispersion were diluted by mixing with equal volumes of NaCl solutions of various concentrations.  This resulted in dispersions with an initial particle volume fraction of $\phi = 0.16$, and electrolyte concentrations of 33--213 m$M$.  These samples were dried in 150 $\mu$m thick Hele-Shaw cells, built from $25 \times 75 $ mm$^2$ glass microscope slides and plastic spacers, held together by clips.  To each cell 175 $\mu$l of dispersion was added, which took about a day to dry into a solid deposit about 15 mm thick, at ambient temperature ($\sim 20^\circ$C) and relative humidity.  

The dried films were imaged in a polarising microscope, between crossed polarising filters, and a half-wavelength filter (first-order retardation plate), as sketched in Fig. \ref{shearbands}(a). In this setup, using white light, birefringence in the film appears as variations in colour -- see Fig. \ref{shearbands}(b,c).  We noticed that as the salt concentration of the dispersion was increased, the intensity of the colour variations decreased; the films appeared more uniform in hue.  Further, for high salt concentrations ($\sim 100$ m$M$), in addition to the main shear bands, many additional fainter shear bands could also be seen, as in Fig. \ref{shearbands}(c).  Finally, at salt concentrations of 143 m$M$ and above, no bands were seen (other than, occasionally, a few bands near boundaries).    

\begin{figure}[]
\centering\includegraphics[width=135 mm]{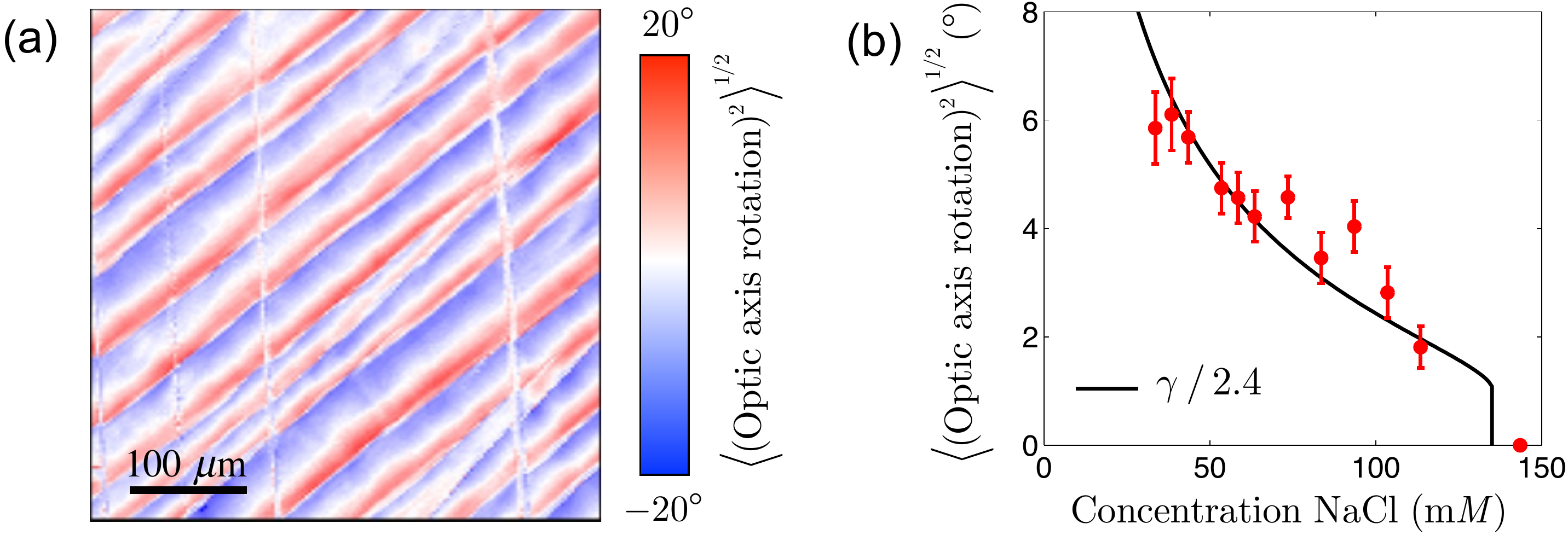}
\caption{Shear bands can reorient the anisotropy of a colloidal film, and can be controlled by adding salt. (a) The pattern of their distortion can be mapped by polarisation microscopy, showing exactly how the bands localise shear strains.  A sample with 43.5 m$M$ NaCl is shown here.  (b) The (root-mean-squared) average changes induced by the bands were measured for dispersions with various initial concentrations of NaCl.  The average reorientations of the film (red points) are proportional to the total amount of uniaxial compression applied to the film, between the gelation and aggregation fronts, as predicted by a DLVO calculation (black line).}
\label{shearbands2}
\end{figure}

To map the distortion caused by the shear bands we rotated the sample stage, collecting images of each film at $10^\circ$ intervals.  The setup was as described above, but in this case the light was passed through a 533 nm filter, before the first polariser.   The resulting images were then digitally counter-rotated so that, by comparing an image series, we could measure the intensity $I$ of the transmitted light through any particular point in the film, as it was turned about an angle $\theta$.  For the green light used, $I(\theta)$ will be minimised when the optic axis is oriented along one of the crossed polarisers.  By fitting a sinusoidal variation in the light, $I(\theta) = I_0 \sin^2\big(2(\theta + \psi)\big) + I_{\rm{bkg}}$, at each pixel, we could measure the orientation $\psi$ of the optic axis across the film, as shown in Fig. \ref{shearbands2}(a).  The average shear that is taken up by the shear bands can thus be related to the root-mean-squared average of the reorientation of the optic axis, or $\langle|\psi|^2\rangle^{1/2}$.  As shown in Fig. \ref{shearbands2}(b), the average twist in the film slowly decreases from about $6^\circ$ for the drying of dispersion at an as-supplied salt concentration, to  $2^\circ$ at about 100 m$M$ salt, just before the bands disappear.

If the shear bands form at the liquid-solid transition, then we can predict the amount of shear available for the bands to release and compare it to what is observed.  As described in Section II.(a), drag forces across this transition provide a compressive force on the colloidal dispersion, which responds by increasing in volume fraction.  Once the particles have formed a soft repulsive solid, they can carry a shear stress, or an anisotropic strain.  To calculate the amount of shear strain available to the shear bands, we assume that the dispersion is compressed uniaxially from the critical volume fraction $\phi_c$, where it first forms a soft solid (i.e. the first points in Fig. \ref{saxs_scaled} where $\gamma>0$), to its final packing fraction $\phi_f$.  Since the material cannot expand in any other direction, the compressive strain that is generated by this process is simply related to the volumetric strain, $\epsilon_x = (\phi_f - \phi_c)/\phi_c$.  This is equivalent to a shear strain of $\gamma = \epsilon_x/2$ at $\pm45^\circ$ to the direction of compression -- the directions along which the shear bands form.  To determine $\gamma$, we thus need to know the gelling concentration, $\phi_c$, of the particles, and its final $\phi_f$.  

For the large Levasil silica particles, and generally for the high salt concentrations required to control the shear banding instability, the osmotic compression experiments shown in Fig. \ref{fig_pressure} disagree with the Poisson-Boltzmann Cell model used in the earlier sections of this paper.  Thus we instead use here a linearised DLVO pair potential of the colloidal silica particles at various salt concentrations and volume fractions, 
\begin{equation}
U = -\frac{Aa}{12s} + \frac{8akT}{L_B}\tanh^2(\varphi_0)e^{-\kappa s}
\label{DLVO}
\end{equation} 
as in \cite{Russel1989,Goehring2010,Boulogne2014,Kiatkirakajorn2015}.  In Eq. \ref{DLVO} $\varphi_0, \kappa^{-1},$ and $L_B$ are the reduced surface electrostatic potential, the Debye length and the Bjerrum legnth, as defined in Section II.(b), while $kT$ is the thermal energy, $a = 46$ nm is the average radius of the particles, and $A = 8\times 10^{-21} J$ is the Hamaker constant for silica \cite{Russel1989}.  The reduced surface potential $\varphi_0$ is calculated as in \cite{Goehring2010,Kiatkirakajorn2015}, assuming a reduced surface charge of 0.16 $e$/nm$^2$ (which matches the zeta potential measurements in \cite{Healy2006}).     For various colloids (silica and polystyrene) it was shown in \cite{Goehring2010} that the particles will gel into a soft repulsive solid when $U$ reaches a few times $kT$, while in \cite{Boulogne2014} the structural anisotropy associated with similar drying colloids was shown to begin at the same point.  Finally, in \cite{Kiatkirakajorn2015} the presence of shear bands in a dried film was shown to require a pair potential of $\sim 5 kT$, using the same potential as Eq. \ref{DLVO}.  Using that approximation we then defined $\phi_c$ as the concentration where the pair potential between two neighbouring particles reached 5 $kT$, and assumed that $\phi_f = 0.64$, or that the final aggregated state is one of random close-packed particles.  

In Fig. \ref{shearbands2}(b) we compare the accumulated shear strain $\gamma$ following from this series of approximations (and expressed as an engineering strain, in degrees), with the average reorientation, $\phi$, observed in dried Levasil films, for different salt concentrations.   We fit $\gamma$ to the data by allowing for a single scaling factor in the magnitude of $\gamma$, of order one, and find that there is good agreement between the strain that is generated across the liquid-solid transition, and the strain released by the formation of shear bands.  The disappearance of the bands at higher salt concentrations, as for the dispersions studied in Ref. \cite{Kiatkirakajorn2015}, is also well-captured by this model.  In other words, we found that the total distortion caused by the shear bands is simply proportional to the total compression of the liquid-solid transition in the drying dispersion.

\subsection{Guiding cracks}

As they dry, many colloidal dispersions also crack, due to capillary forces \cite{Man2008}.  This is a concern for coatings such as paints, and also presents a limit to the manufacture of photonic materials \cite{Zhang2009,Juillerat2006}. However, the control and guidance of cracks in thin films, for example in microfabrication applications, has recently become a topic of some interest \cite{Nam2012,Kim2013,Seghir2015,Nandakishore2016}.  One can imagine such features allowing for the directional control of the friction, conductivity, or permeability of a surface coating, for example.

For a drying colloidal film the direction of crack growth is usually noticed to be perpendicular to the drying fronts (e.g.\cite{Allain1995,Hull1999,Dufresne2003,Pauchard2003,Goehring2010}), at least in simple geometries such as the flat liquid-solid transitions sketched in Fig. \ref{saxs_methods}.  This involves growing from a region of high stress (near the edge of the cell, where evaporation is occurring), to one of low stress (by the liquid-solid transition). Fracture mechanics, however, requires that a growing crack tries to maximise the difference between its local strain energy release rate, and the cost of creating new crack surfaces (the \textit{critical} strain energy release rate) \cite{Lawn1993}. It is thus somewhat surprising that cracks are not more often deflected back towards the edge of the drying layer, where the strain energies are highest.  Here we argue that, instead, cracks in dried colloidal materials are guided by the structural anisotropy of the dried film, and hence the memory of the way in which they dried.  If the packing of particles, and particle-particle contacts, is direction-dependant, then the energy consumed in extending a crack should also be direction-dependant.  All else being equal, the crack will then preferentially grow along the 'easiest' direction (i.e. where the critical strain energy release rate is smallest), just as in crystalline materials a crack will often tend to grow along certain preferred crystal planes.  

\begin{figure}[]
\centering\includegraphics[width=135 mm]{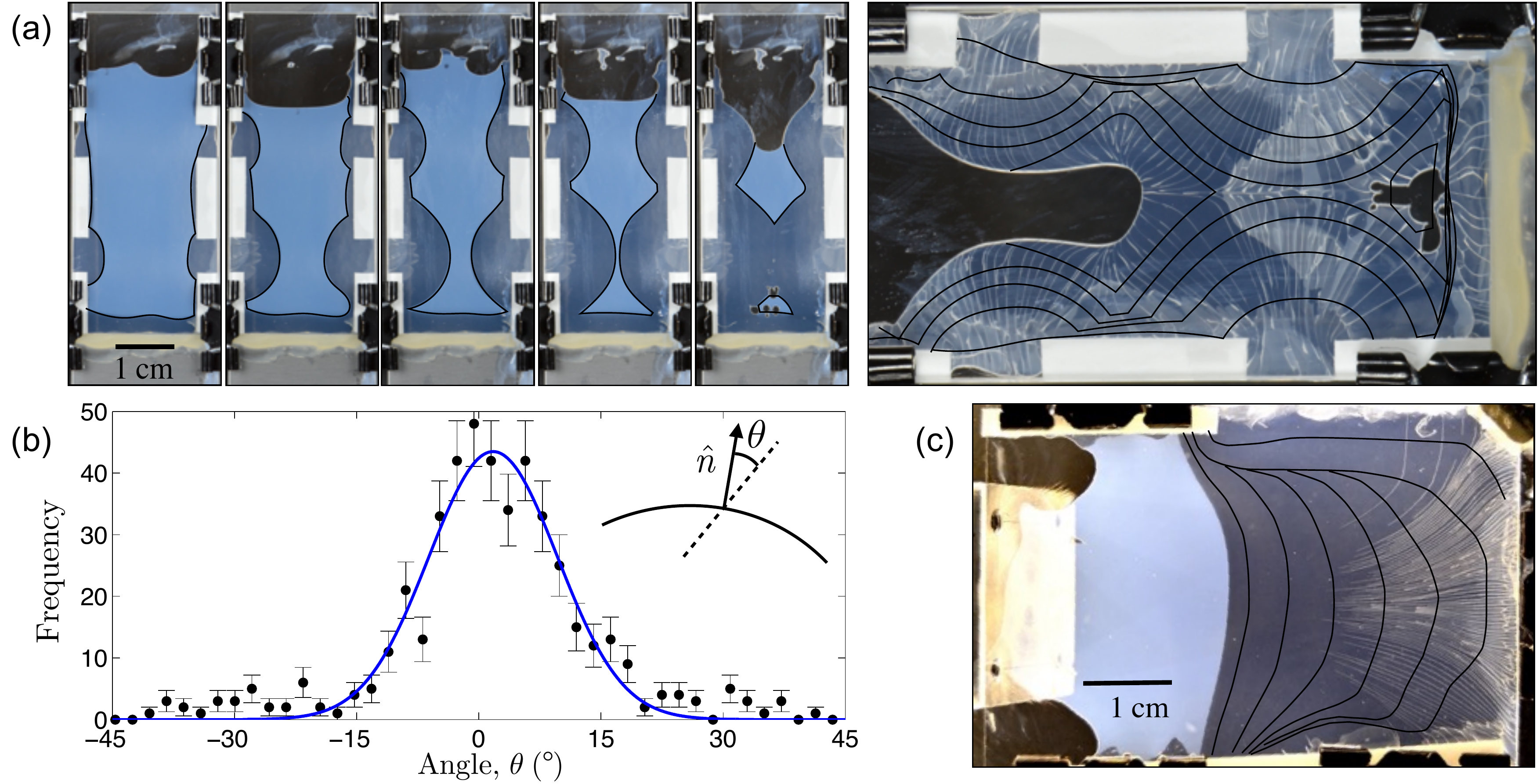}
\caption{Guiding cracks in drying colloidal dispersions.  (a) Colloidal silica (Levasil 30) is dried in a Hele-Shaw cell with a set of small openings on the sides to allow evaporation.  Subsequent images show how drying proceeds in 1-hour steps.  The initially liquid dispersion (milky blue-white) dries into a darker solid deposit (translucent, on black background).  The larger right-hand panel shows how cracks, which appear at the end of drying, lie perpendicular to the superimposed drying profiles (black lines).  (b) The angle between a crack and the outward-pointing normal of the drying front profile, when it had been at that same point, shows a distribution that is sharply peaked around 0 (mean of 463 measurements is $1\pm 1^\circ$, and standard deviation $13\pm 1^\circ$). Panel (c) shows another experiment, where one side of the cell was sealed after 5 hours.  In this case the front positions are shown in 4-hour steps, and one can see how the cracks bend to follow the memory of the drying.  Videos of the drying process for both experiments are provided as supplemental online information.  }
\label{cracks}
\end{figure}

To demonstrate this we again dried Levasil 30 in Hele-Shaw cells.  The appearance of cracks in this dispersion is relatively delayed, and in some cases, as shown in Fig. \ref{cracks}, will not take place until the entire dispersion has fully solidified in the drying cell.  By modifying the pattern of evaporation during drying we could thus guide the drying fronts in relatively arbitrary ways.  Here, as in the shear-band experiments above, we prepared cells from two glass microscope slides (either 75 $\times$ 25 mm$^2$ or 75 $\times$ 50 mm$^2$).   However, various arrangements of spacers and gaps were made around the edges of the cells, to allow for evaporation along different parts of the cell perimeter.  For example, in the experiment shown in Fig. \ref{cracks}(a) there are four small gaps of 5-10 mm along the sides of the cell, and both the top and bottom of the cell have also been left open.  Changes to drying could also be made during an experiment, by sealing any edge with vacuum grease (Dow Corning silicon grease).    At the start of any experiment aqueous dispersions of Levasil 30 (as-received) were pipetted into the cells, which were initially inclined slightly to allow the dispersion to settle to one side.  After a solid layer of material had appeared around the edges of the cell, they were then hung vertically, and time lapse photographs (typically taken every 150 s) were taken of the drying process.  The cells were refilled intermittently during drying, by pipetting additional dispersion into the top edge.  

In each case we found that when cracks form, they are preferentially aligned with the direction along which the dispersion had solidified.  This was true for points throughout the film, even if cracks appeared hours after that region had  aggregated [Fig. \ref{cracks}(a,b)] or if the drying front had moved well on, and had subsequently changed shape [Fig. \ref{cracks}(c)].  For one experiment, Fig. \ref{cracks}(a) shows the drying cell at one-hour intervals, with the position of the liquid-solid front traced out in black.  The final pattern of cracks, which occur after the entire film has solidified, clearly reflects the pattern of drying, and in particular the cracks are parallel to the direction of compression of the material everywhere in the dry deposit.  To confirm this effect quantitatively we measured 463 intersections between cracks and the set of drying front profiles displayed in Fig. \ref{cracks}(a), and determined the misalignment between the outward-pointing normal to the liquid-solid transition at those points, and the direction that a crack subsequently followed.  As shown in Fig. \ref{cracks}(b), these two directions are, on average, very well aligned.  Their difference is fit by a simple Gaussian distribution with a mean of $1\pm 1^\circ$ and a standard deviation of $13\pm 1^\circ$.  In other words, the anisotropy in the film, formed in response to the drag forces of water flowing across the liquid-solid transition region during drying, is generally found to share the same orientation as the subsequent cracking.  We suggest that the change in microstructure of the material changes its fracture energy along different orientations, thus explaining the observed correlation.  

The observations we describe here are similar to the memory effect studied by Nakahara and coauthors (e.g. \cite{Nakahara2006,Nakayama2013,Kitsunezaki2016}), and which was shown in Fig. \ref{fig_intro}(c).  They showed how a wide variety of cues, such as vibration, flow, or standing Faraday waves, can be used to template crack patterns in pastes and slurries.  The common condition for all of their work is that the memory of some event can influence later cracking in materials with a yield-stress, by pre-conditioning the material with an anisotropic pre-stress or strain.  This memory effect appears to also hold true for dried colloidal materials, and to reflect the yield-strain phase that the material temporarily passes through as it changes from a liquid dispersion, to a solid aggregated deposit.

\section{Summary and Conclusions}

As they dry colloidal materials can go through a series of mechanical instabilities including shear band formation, wrinkling, buckling, cracking, delamination, etc.  These responses are controlled by forces that arise from microscopic interactions, between nearby particles and between particles and the fluid that surrounds them.  In order to be able to control these instabilities, one must first understand these interactions, and how they scale up to cause a macroscopic effect.  

We presented an advection-diffusion model of a drying colloidal dispersion in a regular channel.  This one-dimensional representation sought to test when a simple mean-field approach to particle interactions was valid, and when additional details would need to be considered.  The model was fed by a Poisson-Boltzmann Cell (PBC) model of the electrostatic interactions between particles.  This was developed in such a way that it could predict the osmotic pressure and concentration (or collective) diffusivity of a charged colloidal dispersion, and how that dispersion would behave as it was dried.  It had no free parameters, once the size and charge of the colloidal particles was chosen, along with the salt concentration of the dispersant liquid.

The predictions of this pair of models were tested against observations of charged colloidal silica nanoparticles, consisting of three different grades of Ludox dialysed against a variety of salt solutions.  We found that the PBC model accurately predicts the osmotic pressure of these dispersions as they are slowly concentrated to intermediate volume fractions, but that some discrepancies arose at higher salt concentrations (10-50 mM), where the model systematically under-predicted the osmotic pressures of the dispersions.   

Drying experiments were then conducted in Hele-Shaw cells, where small-angle x-ray scattering (SAXS) techniques were used to measure how the particle volume fraction changes across the liquid-solid transitions of directionally-drying colloidal dispersions.  We found that the numerical model of the front correctly captured much of the experimental detail, such as (i) the general shape of the drying front, especially for the smaller particles, (ii) the fact that the concentration profiles across the liquid-solid transition were stretched to be bout an order of magnitude wider than would be expected for particles with only hard-sphere interactions, and (iii) that this stretching of the front was stronger for larger particles.  However, many of the fine details of the concentration profiles were missed.  Most noticeably, when the drying profiles were used to infer an effective diffusivity of the various dispersions, it was found that the colloidal particles showed a marked increase in their collective diffusivity at intermediate-to-high volume fractions, where they were behaving as a yield-stress material, like a paste or gel, rather than a simple fluid.  This increase was not captured by the model, and may represent non-DLVO interactions, or non-isotropic interactions.

As these colloids dry, they undergo a compression along the direction of drying.  We explored the macroscopic implications of these forces in the latter part of this paper, looking at shear bands and cracks.  The shear bands release the uniaxial compression of the film by allowing for slip at $\pm 45^\circ$ to the direction of compression.  In particular, we showed that the amount of slip accommodated by any of these bands was proportional to the total amount of deviatoric strain that would have accumulated across the liquid-solid transition, had it not been relieved by the shear bands.   Both the appearance of shear bands, and the extent of their shear, can thus be controlled by adjusting the chemistry of the starting dispersion, before it dries.     The cracks, in turn, release strain energy by allowing the dispersion to shrink more as it dries.  We demonstrated that the appearance and paths of cracks could be guided by the structural anisotropy that the liquid-solid transition leaves behind.  
 
\enlargethispage{20pt}


\acknowledgements{The authors wish to thank B. Cabane and D. Fairhurst for discussions.  The SAXS experiments were performed on beamline ID02 at the European Synchrotron Radiation Facility (ESRF), Grenoble, France and we are grateful to M. Sztucki at the ESRF for providing assistance.}



\end{document}